\newcommand{\E}{{\rm e}}
\newcommand{\ptd}{\partial}
\newcommand{\lt}{\left}
\newcommand{\rt}{\right}
\newcommand{\no}{\nonumber}
\newcommand{\kt}{\rangle}

\newcommand{\ket}[1]{| #1 \rangle}

\newcommand{\av}[1]{\langle #1 \rangle}
\newcommand{\be}{\begin{equation}}
\newcommand{\ee}{\end{equation}}
\newcommand{\bea}{\begin{eqnarray}}
\newcommand{\eea}{\end{eqnarray}}
\newcommand{\rf}[1]{(\ref{#1})}

 \documentstyle[aps,eqsecnum,pra]{revtex}

\title{Semiclassical spin damping: Superradiance
revisited}
 
\author{Petr A.~Braun}
 \address{Department of Theoretical Physics,
  Institute of Physics,
  Saint-Petersburg University,
  Saint-Petersburg, 198904 Russia}

\author{Daniel Braun, Fritz Haake, Joachim Weber}
\address{Fachbereich Physik,
  Universit\"at-Gesamthochschule Essen,
  45117 Essen, Germany}

\begin{document}
\maketitle
\begin{abstract}
A well known description of superradiance from pointlike collections of many
atoms involves the dissipative motion of a large spin. The pertinent
``superradiance master equation'' allows for a formally exact solution which
we subject to a semiclassical evaluation. The clue is a saddle-point
approximation for an inverse Laplace transform. All previous approximate
treatments, disparate as they may appear, are encompassed in our systematic
formulation. A byproduct is a hitherto unknown rigorous relation between
coherences and probabilities. Our results allow for generalizations to spin
dynamics with chaos in the classical limit. \\

\end{abstract}

\newpage
\section{Introduction}

Dissipative motion of large spins was first seen in experiments on
superradiance or superfluorescence (For extensive reviews see Refs.
\cite{Haroche,Benedict}), after being proposed a lot earlier by Dicke
\cite{Dicke}. The so called superradiance master equation proposed in
\cite{BSH,Agarwal} has since become a standard tool for describing
the collective dynamics of identical superradiating atoms in the
small-sample limit. Formally speaking, it provides a quantum treatment
of a large spin with conserved square, ${\bf J}^2=j(j+1)$, with the
quantum number $j$ capable of taking on positive half integer or integer
values up to half the number of atoms $N$. The origin of such an angular
momentum lies in the familiar formal equivalence of a single two-level
atom to a spin-$\frac{1}{2}$. In (semi)classical parlance, the spin in
question is called the Bloch vector whose $z$-component measures the
energy stored in atomic excitation while the transverse components are
related to the dipole element responsible for the atomic transition.
More or less everything worth knowing about the superradiance master
equation in relation to the numerous superfluorescence experiments
has been worked out more than a decade ago.

When we pick up the thread now our motivation is not to better explain
anything previously observed, but rather  the expectation of new
experiments involving dissipative motion of large spins constituted by
many identical two-level atoms, albeit motions that would have a chaotic
classical limit and display quantum manifestations of chaos when the
spin quantum number $j$ is of the order of several hundreds or
thousands. When beginning to look into such dynamics \cite{qsc} we
found, somewhat to our surprise, that previous treatments of the
superradiance master equation were so directly geared to the specifics
of superradiant pulses as transient events that new questions do indeed
require some new theoretical work. In particular, the semiclassical
limit of large $j$ deserves systematic attention and turns out to harbor
one or the other surprise which we begin to uncover in the present
paper.

The large-$j$ limit can be approached through the rigorous solution of
the master equation which was known from the very beginning
\cite{BSH}, and we shall actually follow that path here. Strangely
enough, up to now that rigorous solution has mostly been looked upon as
a curiosity rather than a useful starting point of analytic work; even
numerical evaluations were disfavored against routines for solving
coupled differential equations for density matrix elements in some
representation.

We propose to show that the large-$j$ limit is very conveniently
accessed by subjecting the rigorous Laplace transformed density matrix
to a saddle-point evaluation of the inverse Laplace transformation. More
specifically, we carry out this program in the eigenrepresentation of
$J_z$ and ${\bf J}^2$ for the density matrix $\langle
jm|\rho(t)|jm'\rangle$ and the propagator relating that density matrix
to its initial form $\langle jm|\rho(0)|jm'\rangle$. The saddle-point
result turns out reliable provided that not only $j$ is large but also
the difference between the initial and final eigenvalues of $J_z$, i.e.
$|m-m'|\gg 1$. That restriction unfortunately affects the propagator at
early times while most of the probability still resides in levels $m$
close to the initial $m'$. We therefore establish an independent
early-time propagator, show its agreement with the saddle-point version
in a certain time span and finally combine the two to an explicit
expression of uniform validity.

Our uniform propagator turns out to systematically encompass previous
asymptotic results. Among these is, trivially, the fully classical
behavior arising in the limit $j\rightarrow\infty$ as long as the
initial state is not too close to the state of full initial excitation
$m=j$ which in the classical limit is an infinitly long-lived state of
marginal equilibrium. The classical behavior in question is that of an
overdamped pendulum. The pertinent equation of motion for the so-called
Bloch angle $\theta$ (defined through
$\cos\theta=\lim_{j\rightarrow\infty}\langle J_z(\tau)\rangle/j$) reads,
with $\tau$ denoting a suitably scaled time,
$\frac{d}{d\tau}\theta=\sin\theta$; the the well known solution is
\be
\tan\frac{\theta(\tau)}{2}={\rm e}^\tau\tan\frac{\theta(0)}{2}\,.
\label{classtraj}
\ee
Furthermore, we recover the random-jitter picture first suggested in
\cite{Giorgio1,Giorgio2} and the ensuing distribution of delay times as
well as the scaling results for time dependent expectation values of
products of the observables $J_x, J_y, J_z$ obtained by somewhat
hit-and-run methods in \cite{HG,GH}.

An interesting byproduct of our investigation is an exact relation
between diagonal and offdiagonal elements of the density matrix in the
$jm$-basis, which to the best of our knowledge has previously gone
unnoticed. One may thus confine all work towards solving the master
equation to the probabilities $\langle jm|\rho(t)|jm\rangle$ and
eventually obtain the coherences $\langle jm|\rho(t)|jm'\rangle$
through the relation in question.

A subsequent paper will deal with the large-$j$ limit with the WKB method.
 
\section{Master equation and dissipative propagator}

The two states of an atom resonantly coupled to a mode of the
electromagnetic field may be thought of as the states of a
spin-$\frac{1}{2}$, and all observables of the effective two-level
atom can be represented as linear combinations of unity and the three
spin operators $J_x, J_y, J_z$. In particular, the energy may be
associated with $J_z$ and the other two spin operators with the
atomic dipole moment. If $N$ such atoms, all identical,
couple collectively to the electric field $E$ one has an interaction
Hamiltonian $\propto -J_xE$ where $J_x=\sum_{\mu=1}^N J_x^{\mu}$ is
the sum of all single-atom contributions; similarly, one has a
global atomic energy $\propto J_z=\sum_{\mu=1}^N J_z^{\mu}$. The
collective spin operators obey the familiar angular-momentum
commutation relations $[J_x,J_y]={\rm i}J_z$ etc. The Hilbert space
for the $N$ atoms is $2^N$ dimensional but falls into subspaces not
connected by the collective observables $J_i$; each subspace has
fixed ${\bf J}^2=j(j+1)$ with nonnegative integer or half-integer $j$
not exceeding $N/2$. The $(2j+1)$ states in the $j$th subspace are
conveniently taken as the eigenstates $|jm\kt$ of $J_z$ with
eigenvalues $m=-j,-j+1,-j+2,\ldots ,j$. The highest energy may be
associated with $m=j$ whereupon the ground state has $m=-j$. In
particular, the subspace with $j=\frac{N}{2}$ consists of $N+1$
states which are all totally symmetric in all atoms; that space may
be singled out experimentally by preparing all atoms in their lower
state.

In the superradiance experiments of Ref. \cite{Haroche2} a single mode of
the electromagnetic field within a resonator was coupled to $N$
two-level atoms such that the dynamics was that of the so-called
Jaynes-Cummings model, with dissipation included to account for field
losses from the resonator.
In the limit of overdamped Rabi oscillations the field mode can be
eliminated adiabatically. A master equation for the atomic density
operator thus results \cite{BSH,Agarwal} of which we shall consider
the low-temperature version, thus forbidding the atoms to pick up
thermal photons from the environment,
\be \frac d {dt}\hat
{\rho}=\kappa\{[J_-,\hat{\rho}J_+]+[J_-\hat{\rho},J_+]\}\,; \label{br1}
\ee
here $J_{\pm}=J_x\pm iJ_y$ are the familiar raising and lowering
operators and $\kappa$ measures the rate of photon loss from the cavity.
 
In the basis set $|jm\kt$ we obtain from \rf{br1} a set of equations for
the elements $\rho_{m_1m_2}=\av {jm_1|\hat{\rho}|jm_2}$ of the density
matrix,
\be
\dot{\rho}_{m_1m_2}=2\kappa\lt[\sqrt{g_{m_1+1}g_{m_2+1}}\rho_{m_1+1,m_2+1}
-\frac{g_{m_1}+g_{m_2}}{2}\rho_{m_1,m_2}\rt]
\label{br2} \ee in which $g_m$ denotes the ``rate function'' \be
g_m=j(j+1)-m(m-1).
\ee
The diagonal element $\rho_{mm}$ of the density
matrix gives the probability to find the system of atoms in the state
$\ket {jm}$; the elements $\rho_{m_1m_2}$ with $m_1-m_2\ne0$ will be
referred to as coherences. It is worth noting a certain
unidirectionality of the flow of probability and coherence, downwards
the $m$-ladder, the physical origin of which is of course the
low-temperature limit mentioned above. A further important feature of
the system (\ref{br2}) is that the density matrix elements with
different $m_1-m_2$ evolve independently. To make that independence
manifest it is convenient to introduce the quantum numbers \be
m=\frac{m_1+m_2}{2},\quad k=\frac{m_1-m_2}{2} \ee which can be
simultaneously either integer or half-integer. Accounting for \be
\frac{g_{m_1}+g_{m_2}}{2}=g_m-k^2 \ee and changing the notation
$\rho_{m_1m_2}$ for the density matrix element to $\rho^k_m$ we can
rewrite the master equation as	 \be
\frac{d\rho_m^k}{dt}=2\kappa\lt[\sqrt{g_{m+k+1}g_{m-k+1}}\rho_{m+1}^k-(g_m-k^2)\rho_m^k\label{master2}\rt].
\ee It is now indeed obvious that the ``skewness'' $k$ enters only as a
parameter.
 
The linear relation between the density matrices at the current time and at
the initial moment, 
\be
\rho_m^k(t)=\sum_n D^k_{mn}(t)\rho_n^k(0)\,,
\label{propadef}
\ee
defines the $k$-dependent matrix $D^k_{mn}(t)$ which will be called the
dissipative propagator. Its column corresponding to a certain fixed $n$
can be regarded as the solution of the master equation (\ref{master2})
corresponding to the initial condition $\rho^k_{m}(0)=\delta_{mn}$. Due
to the unidirectionality of the master equation it is obvious that
$D^k_{mn}=0$ if $m>n$. We shall drop the superscript $k$ in the case
$k=0$, i.e. when the diagonal elements of the density matrix are
considered.

\section{Saddle-point asymptotics of the dissipative propagator}

There have been a number of successful attempts to treat the large-$j$
limit of the superradiance problem
\cite{BSH,HG,GH,Giorgio1,Giorgio2,Haroche}. These were
concerned with the solution of the master equation for certain
particular cases or directly aimed at specific average properties of the
process. The purpose of the present paper is to establish uniform
asymptotics of the dissipative propagator without such restrictions. We
use the exact solution of the master equation in the form of the Laplace
integral which was obtained long ago \cite{BSH} but remained largely
unexplored. Previously established results for the propagator, the
distribution of delay times, and time dependent expectation values follow
from our uniform asymptotic propagator.

Before embarking on our proposed asymptotic adventure it is convenient to
adopt the parameter 
\be
\sqrt{j(j+1)}\approx j+\frac{1}{2}\equiv J
\label{witchcraft}
\ee
as a measure of the ``size'' of the angular momentum; the semiclassical
formulae to be established take a prettier form if we use $J$ rather
than $j$.

\subsection{Laplace representation of the  exact propagator }

Following \cite{BSH} let us recall the Laplace integral
representation of the propagator. Defining the Laplace image as ${\cal
D}_{mn}^k(z)=\int_0^{\infty}\E^{-zt}D^k_{mn}(t)dt$ we turn our master
equation into a recursion relation with the easily found solution
\be
{\cal D}_{mn}^k(z)=\frac{1}{2\kappa
\sqrt{g_{m-k}g_{m+k}}}\prod_{l=m}^n\frac{\sqrt{g_{l-k}g_{l+k}}}
{\frac{z}{2\kappa}+g_l-k^2}\,. 
\ee
To get the dissipative propagator itself we invert the Laplace
transform. Introducing a scaled time
\be
\tau=2\kappa Jt \label{taut}
\ee
and the quantity
\be Q_{mn}=\prod_{l=m+1}^n
g_l=\frac{(j+n)!(j-m)!}{(j+m)!(j-n)!} \label{defQ}
\ee
we bring our propagator to the form
\be D_{mn}^k(\tau)=
\frac{\sqrt{Q_{m-k,n-k}Q_{m+k,n+k}}}{2\pi
i}\int_{b-i\infty}^{b+i\infty}dv\;\E^{\tau
v/J}\prod_{l=m}^n\frac{1}{v+g_l-k^2}\,,\label{prop2}
\ee
where $b$ should be larger than the largest pole in
the denominator.

\subsection{Relation between densities and coherences}

An unexpected new result of the representation \rf{prop2} is an
identity connecting the propagators for the diagonal and for the
off-diagonal elements of the density matrix,
\be
D_{mn}^k(\tau)=D_{mn}(\tau)\frac{\sqrt{Q_{m-k,n-k}Q_{m+k,n+k}}}{Q_{mn}}\E^{
k^2 \tau/J}.\label{identity}
\ee
For the proof it is sufficient to
shift the integration variable in \rf{prop2} to $\bar{v}=v-k^2$.
Alternatively, the connection between the diagonal and off-diagonal
density matrix elements can be checked by entering the master equation
with the ansatz
\be \rho_m^k=\frac {(j+m)!}{(j-m)!} \frac
 {\sqrt{(j-m-k)!(j-m+k)!}}
{\sqrt{(j+m-k)!(j+m+k)!}}
\E^{2\kappa k^2t}\tilde{\rho}_m(t)\,;
\label{identity2}
\ee
the new unknowns $\tilde{\rho}_m(t)$ then turn out to evolve in time like
probabilities, i.e. to obey
(\ref{master2}) for $k=0$.

The positive sign of the exponents in these relations between
probabilities and coherences is not a misprint: the coherence
$\rho_m^k=\rho_{m+k,m-k}$ does decay more slowly than the density
$\rho_m=\rho_{m,m}$. Moreover, there is no conflict with the nowadays
popular phenomenon of accelerated decoherence
\cite{LegCal,Zurek,HaaZuk}: Quantum dissipative processes do imply much
larger decay rates for coherences than for probabilities but only so
with respect to certain states which are distinguished by the process
itself; for the dissipative process studied here such distinguished
states are, for instance, coherent angular-momentum states
\cite{Arecchi,GH} but not the states $|jm\rangle$.

A simple illustration of the statement just made may be helpful, even if
it amounts to sidestepping to another dissipative process for an angular
momentum, the one described by the master equation \cite{qsc}
$\dot{\rho}=\kappa\{[J_z,\rho J_z]+[J_z\rho,J_z]\}$. In that case the
eigenstates $|jm\rangle$ of $J_z$ are the distinguished ones as is
obvious from $\dot{\rho}_m^k=-4\kappa k^2\rho_m^k$ : The probabilities
$\rho_m^0$ are all conserved while the coherences have decay rates
growing quadratically with the skewness k.

\subsection{Saddle-point evaluation of the Laplace integral}

The relation \rf{identity} between probabilities and coherences clearly
allows us to confine the remaining investigation to the case $k=0$, i.e.
to the propagator of the densities. Our goal is to do the integral in
the exact formula \rf{prop2} in the limit of large $J$. To begin with,
let us rewrite that formula for $k=0$ as
\be D_{mn}(\tau)=\frac {Q_{mn}}
{2\pi i}\int_{b-i\infty}^{b+i\infty}\E^{Z_{mn}(v,\tau)}\;dv \label{repr5}
\ee
with the exponent
\be Z_{mn}(v,\tau)=\tau v/J-\sum_{l=m}^{n}\ln
(v+g_l)\,.
\ee
Suppose now that $n-m\gg 1$. Then since the number of
terms in the sum $Z$ is proportional to $n-m$ its value is generally
also large, which fact suggests a saddle-point approximation. The
stationary points of the exponent are given by the solutions for $v$ of
\be
Z_{mn}^{'}=\tau/J-\sum_{l=m}^{n}\frac{1}{v+g_l}=0\,. \label{statp1}
\ee
All roots of this saddle-point equation are real as is immediately seen
by putting $v=x+iy$ and separating the imaginary part. We further note
that to the right of the largest pole $v_{{\rm max}}={\mbox {\small max} \atop
m\le l\le n}\; \{-g_l\}$ of the integrand in (\ref{prop2}) the sum in
(\ref{statp1}) decreases monotonically from $+\infty$ to $0$ as $v$
grows from $v_{{\rm max}}$ to infinity. Therefore we have one and only one root
$v_0$ in that domain. Its position depends on the time $\tau$: When
$\tau$ goes to zero $v_0$ tends to infinity; conversely, for
$\tau\rightarrow\infty$ the saddle point $v_0$ approaches the pole at
$v_{{\rm max}}$.
 
The second derivative with respect to $v$ of the exponent,
 \be
Z_{mn}^{''} =\sum_m^n \frac{1}{[v+g_l]^2}\,,
\ee
is positive for real $v$ which means that the direction of steepest
descent from the saddle is parallel to the imaginary axis. The
saddle-point approximation for the integral \rf{repr5} thus gives \be
D_{mn}\approx \frac{Q_{mn}}{\sqrt{2\pi Z''_{mn}}}\E^{Z_{mn}(v_0,\tau)}\,.
\label{asymp1} \ee

\subsection{Euler-Maclaurin estimates for the sums}

To render the expression \rf{asymp1} useful, we must evaluate the three
sums in $Z_{mn}, Z_{mn}^{'}, Z_{mn}^{''}$. The familiar Euler-Maclaurin
summation formula $\sum_m^n f(k)\approx\int_m^n f(x)dx+(f(m)+f(n))/2$
comes to mind first but is not immediately suitable for our purpose. We
rather employ a modified version which involves nothing but an integral;
to compensate for the absence of the extra boundary terms the
integration interval is extended,
\be
\sum_m^n f(k)\approx\int_{m-1/2}^{n+1/2} f(x)dx.
\label{our}
\ee
The accuracy of both summation rules is the same for smooth summands $f(k)$.

In applying \rf{our} to the sum in the saddle-point equation we rewrite
the rate function as $g_l=J^2-(l-1/2)^2$, introduce the rescaled
variables
\be
\mu=\frac{m-1}{J},\quad \nu=\frac{n}{J},\quad a=\frac{\sqrt{v_0+J^2}}{J}
\label{convention}
\ee
and obtain
\be
\sum_{l=m}^{n}\frac{1}{v+g_l}\approx \int _{m-1}^n \frac {dx}{v_0+J^2-x^2}
=\frac{1}{2Ja}\ln \lt[\frac{(a+\nu)(a-\mu)}{(a-\nu)(a+\mu)}\rt]\,.
\ee
The saddle-point condition \rf{statp1} thus takes the form
 \be 
\tau ={1 \over 2a} \ln  { \left( a+\nu  \right)  \left( a-\mu  \right)
 \over  \left( a-\nu  \right)  \left( a+\mu  \right) }\,.
\label{pab131} 
\ee 
It determines $a$ as a function of $\nu,\mu$, and $\tau$. As already
explained above, the single root of interest is positive and larger than
the larger of $|\mu|,|\nu|$.
 
Similary proceeding with the sums in $Z_{mn}^{''}$ and $Z_{mn}$ we find 
\bea
 J^3 Z_{mn}^{''}&=& \lt.\frac 1 {2a^2}\lt(\tau+\frac {\nu}{a^2-\nu^2}
 -\frac {\mu}{a^2-\mu^2}\rt)\rt|_{a=a(\mu,\nu,\tau)}\equiv\Xi(\mu,\nu,\tau)\,,
\label{defxi2}\\
Z_{mn}(v,\tau)&=& J\lt[\tau(a^2-1)-2(\nu-\mu)\ln J+2(\nu-\mu)
- \sigma(a,\mu,\nu) \rt]
\label{s0}
\eea
with the auxiliary function
 \bea
 \sigma(a,\mu,\nu)&\equiv& (\nu+a)\ln(\nu+a)-(\mu+a)\ln(\mu+a)\no\\
&-&(a-\nu)\ln(a-\nu)+(a-\mu)\ln(a-\mu)\,.\label{sigma}
\eea

We should comment on the slight asymmetry in the definitions of the
macroscopic variables $\mu$ and $\nu$ in \rf{convention}. The use of
$(m-1)/J$ instead of $m/J$ as the macroscopic variable $\mu$ is formally
related to our extension by 1 of the integration interval in the
summation formula \rf{our} and has the benefit of preventing the small
parameter $1/J$ from appearing explicitly in the saddle-point equation
\rf{pab131}.

\section{Uniform  asymptotics of the propagator}

We came to our saddle-point approximation assuming that the number of
terms in the sum $Z_{mn}$ equal to $n-m$ is large. It is not surprising
therefore that the approximation \rf{asymp1} loses its accuracy when
$n-m$ is of the order unity or zero; that situation prevails, e.g., for
small times $\tau$; an alternative approximation is then desirable and
will be constructed presently.

\subsection{Small-time approximation}

To explain the essence of the new approximation let us give a simple
example. Consider the Laplace image function with two simple poles
${\cal V}(z)=(z-c-d)^{-1}(z-c+d)^{-1} $ and its original function
$V(t)=\E^{ct}\;d^{-1}\sinh td$. As long as $ td\ll 1$ the hyperbolic sine
can be replaced by its argument such that $V(t)\approx t\E^{ct}$. We have
thus in effect replaced the two close by poles of the Laplace image by a
single second-order pole; that replacement is obviously justified for
sufficiently small times.
 
To employ this observation for the Laplace representation of the propagator
\rf{prop2} we introduce the new integration variable  $x=\tau v/J$ and obtain  
\be
D_{mn}(\tau)=Q_ {nm} \lt(\frac{\tau}{J}\rt)^{n-m}\frac{1}{2\pi i}
\int^ {b+i\infty}_{b-i\infty}\frac{\E^x dx}{\prod_{l=m}^n[x+g_l\tau/J]}
\,.\label{int2}
\ee
The length of the interval on which the poles of the integrand now lie is
proportional to $\tau$, 
\be
|g_m-g_n|\,\frac{\tau}{J}=\frac{|m+n-1|}{J}\,(n-m)\tau\,.
\label{param}
\ee
If that length is much smaller than unity the poles of the integrand  of
(\ref{int2}) are nearly degenerate, and that proximity enables us to replace
the product in 
the denominator by the $(n-m)$-th power of the average factor
$x+\bar{g}\tau/J$ with
$\bar{g}\equiv g_{\frac{m+n}{2}}=J^2-\lt(\frac{n+m-1}{2}\rt)^2$.
The integral is then easily calculated and yields the small-time
asymptotics of the dissipative propagator,
\be
D_{mn}(\tau)=\frac{Q_ {mn} }{(n-m)!}\lt(\frac{\tau}{J}\rt)^{n-m}
\exp{\left\{\d-\frac{\tau}{J}\lt[J^2-\lt(\frac{n+m-1}{2}\rt)^2\rt]
\right\}}\,.\label{smalltau1}
\ee
Unlike the saddle-point approximation, the foregoing expression is fully
explicit. We shall keep referring to it as the small-time approximant
although the underlying small parameter is the combination \rf{param} of
both $\tau$ and the quantum numbers $m,n$.

\subsection{Matching the two approximations}
 
The saddle-point and the small-time approximations for the propagator
practically coincide for an intermediate range of arguments. Let us assume
$l= n-m+1\gg 1$ but on the other hand
$\zeta\equiv l/J\ll 1$ ( say, $l\sim \sqrt{J}$). The solution
$a(\tau, \mu, \nu)$ of the saddle-point equation \rf{pab131}  can then
be found by expanding  in powers of $\zeta$,
\be
a^2=\nu^2+\frac{\zeta}{\tau}-\zeta\nu+{\cal O}(\zeta^2)\,.\label{devel}
\ee
The exponent (\ref{s0}) in the saddle-point formula then simplifies
according to $\tau(a^2-1) \approx \zeta+\tau\lt[(\nu-\zeta/2)^2-1\rt]$
and $\sigma(a,\mu,\nu) \approx \zeta\lt(\ln \frac{\zeta}{\tau}+2\rt)$
while the prefactor becomes $\Xi\approx \tau^2/\zeta$. Collecting these
pieces in \rf{asymp1} we obtain 
\be
D_{mn}=\lt.\frac {Q_{mn}\sqrt{l}}
{\sqrt{2\pi}}
\lt(\frac{\E }{l}\rt)^l
\lt(\frac{\tau}{J}\rt)^{l-1}
\E^{-\frac{\tau}{J}
\lt[J^2-\lt(n-\frac{l}{2}\rt)^2\rt]}\rt|_{l=n-m+1}.
\label{expa}
\ee
This in turn is the small-time approximation \rf{smalltau1} provided we
there replace the factorial $(n-m)!=(l-1)!$ \`a la Stirling,
$(l-1)!\approx \sqrt{\frac{2\pi} {l}}\lt(\frac l \E \rt)^l$. Hence the
saddle-point and small-time approximations agree for $1\ll l \ll J$.

\subsection{Uniform approximation}

The two approximations under discussion can be merged into a single  one
which generally behaves like the saddle-point formula \rf{asymp1} but
preserves its accuracy even when $m$ is close to $n$ and/or the time
$\tau$ is small. We just have to divide the saddle-point result
\rf{asymp1} by the ratio of the  factorial $(n-m)!$ to its Stirling
approximant. If $n-m$ is large that ratio is unity
but otherwise the correction replaces the saddle-point
version with the small-time propagator
\rf{smalltau1}. We thus obtain the principal result of our paper for the
density propagator in the large-$j$ limit, 
\bea
D_{mn}= \frac{Q_{mn}J^{3/2}}{(l-1)!\sqrt{l \;\Xi \;}}
\lt(\frac {l \E}{J^2} \rt)^l  
\E^{ J\lt[\tau(a^2-1)
- \sigma(a,\mu,\nu) \rt]
   },\label{uniform}\\
l=n-m+1, \quad\mu=(m-1)/J,\quad\nu=n/J,\quad a=a(\mu,\nu,\tau)\,.\no
\eea
It is valid in a wide range of quantum numbers and propagation times and
thus merits the name uniformly asymptotic propagator. The error is of
order $1/J^2$ except for the not very interesting late times when the
bulk of the probability has settled in the lowest level; that latter
restriction for $\tau$ arises due to the close encounter of saddle and
pole mentioned in Sect. IIIC.
 
We have checked that \rf{uniform} provides an efficient tool to
numerically calculate the dissipative propagator; if $j$ is large its
accuracy becomes comparable or even superior to that of the numerical
integration of the master equation. The only inconvenience is the
necessity to determine the saddle-point parameter $a=a(\mu,\nu,\tau)$ by
solving \rf{pab131} which generally has to be done numerically.

\section{Special cases}
We proceed to considering situations in which the uniform approximation
simplifies. The strategy invariably is to approximate factorials of
large numbers \`a la Stirling. Some cases even allow for an analytical
solution for the saddle-point parameter $a$ whereupon fully explicit
formulas for the propagator arise. Some well-known results of
superradiance theory are thus recovered and revealed as special cases of
the uniform approximation.

 \subsection{Semiclassical approximation}
The uniformly asymptotic propagator \rf{uniform} depends on the quantum
numbers $m,n,j$ in two ways. First there is the factorial dependence
which reflects the discrete character of the representation. Second,
there is the dependence on the arguments $\mu,\nu$ which can be regarded
as the classical counterparts of $m,n$ scaled with respect to the total
angular momentum; they tend to continuous variables in the classical
limit.

Suppose we are not interested in effects tied up with the discreteness
of quantum levels and want to obtain a smooth function of the
macroscopic coordinates $\mu,\nu$ only. This is easily achieved by
replacing the factorials $(n-m)!, \; (j\pm n)!,\;(j\pm m)!$ by their
Stirling estimates. While such a replacement would be unacceptably
inaccurate if the arguments $m,n$ approached $\pm j$ (``the poles'' of the
Bloch sphere in classical parlance) or each other, it otherwise
reliably yields
\bea
D_{mn}(\tau)&=&\frac 1 {(1-\mu^2)\sqrt{2\pi J\Xi\;}}\E^{J\Phi(\mu,\nu,
\tau)},\label{asymd2}\\
\Phi(\mu,\nu,\tau)&=&\tau(a^2-1)-\sigma(a,\mu,\nu)+\sigma(1,\mu,\nu)\,.
\eea
We here speak of the semiclassical approximation because of the implied
assumption that all the quantum numbers and their relevant combinations
are large. As a function of $\mu$ at fixed $\nu$ and $\tau$ the
semiclassical propagator displays a single maximum located according to
\be
\frac {\ptd \Phi}{\ptd \mu}=\ln \frac {a^2-\mu^2}{1-\mu^2}=0
\label{fomca}\,,
\ee
i.e. $a=1$. The saddle-point equation \rf{pab131} then yields the most
probable value of $\mu=J_z/J$ at time $\tau$ related to the initial value
$\nu$ through
\begin{equation} 
\tau ={1 \over 2} \ln  { \left( 1+\nu  \right)  \left( 1-\mu  \right) 
 \over  \left( 1-\nu  \right)  \left( 1+\mu  \right) }\,. \label{pab1311}
\end{equation}  
Written in terms of the polar angle of the Bloch vector
$\cos\Theta=\mu,\;\cos\Theta_0=\nu$ the last equation becomes the solution
of the equation of motion of the overdamped pendulum (\ref{classtraj})
mentioned in the Introduction. Indeed, the 
classical picture of the atomic dynamics in superradiance is that of the
Bloch vector creeping from whatever initial orientation $\theta_0$
towards the equilibrium $\theta=\pi$ like an overdamped pendulum with
the azimutal angle $\phi=\arctan(J_x/J_y)$ fixed. None too surprisingly,
the maximum of the distribution $D_{mn}$ with respect to $m$ occurs at
the point $m=m(n,\tau)$ predicted by the classical motion of the Bloch
vector.

As it stands in \rf{asymd2} the semiclassical propagator correctly
describes a broadening of the initially sharp distribution
$D_{mn}(\tau=0)=\delta_{mn}$ to one with a width $\propto\sqrt{J}$. For
many applications that width is negligible such that we may replace the
propagator by
\be
\lim_{J\rightarrow\infty}J D_{mn}(\tau)=
 \delta(\mu-\mu(\tau,\nu))\label{asymd3}
\ee
where $\mu(\tau,\nu)$ is the classical trajectory according to \rf{pab1311}. 
For instance, expectation values like  $\langle J_+^sJ_z^kJ_-^{s}\rangle$
can be calculated to leading order in $J$ with the help of the
foregoing sharp version of the semiclassical propagator through the integrals
\be
\langle J_+^sJ_z^kJ_-^{s}\rangle=
J^{2s+k}\int_{-1}^1 \delta(\mu-\mu(\tau,\nu))(1-\mu^2)^s \mu^kd\mu
= (1-\mu(\tau,\nu)^2)^s \mu(\tau,\nu)^k\,,
\ee
provided, it is well to repeat, the initial point $n=J\nu$ is well
removed from the most highly excited ones, $j-n\gg 1$. No quantum
effects at all survive in that expression; they would only show up as
small standard deviations at most of order $1/\sqrt{J}$ if the small
width of the propagator (\ref{asymd2}) were kept.

\subsection{Early stage of superradiant decay of highest-energy
initial states  \label{up}}

We now take up the previously best studied aspect of superradiance, the
decay of the most highly excited atomic initial states, $j-n \ll j$. We
begin by studying the early stage, i.e. small $\tau$, while the bulk of
the probability still resides with highly excited states. This means
that only those propagator elements are significantly different from
zero for which the final quantum number $m$ is also close to $j$, or
$j-m\ll j$.

We are so led to examine our uniform approximation when the macroscopic
variables  $\nu$ and $\mu$ are close to unity. Expanding the solution of
the saddle-point equation \rf{pab131} in powers of $1-\nu,1-\mu$ we find
the function $a(\mu,\nu,\tau)$ in terms of the nonlinearly rescaled time
\be
\xi=\E^{-2\tau}\ee
as
$a\approx \frac{\nu-\mu \xi}{1-\xi}$.
From here it is easy to establish the ingredients of the uniform
propagator \rf{uniform},
\bea
(a^2-1)\tau-\sigma&\approx& (1-\nu)\ln\xi+(\nu-\mu)\lt[\ln(1-\xi)
-\ln(\nu-\mu)-\ln 2\mbox{\rm e}\rt],\no\\
\Xi&\approx &
\frac{\mbox{sinh}^2\;\tau}{\nu-\mu},\quad Q_{mn}
\approx (2J)^{n-m}\frac{(j-m)!}{(j-n)!}\,,
\eea
which bring the propagator to the limiting form  
\be
D_{mn}(\tau)={j-m \choose j-n} \xi^{j-n+1}(1-\xi)^{n-m}\,, \label{harmnorth}
\ee
known as the linear approximation describing the early stages of the
superradiant process \cite{Haroche}.

\subsection{Bright stage of superradiant decay of highly excited initial
states}  

Suppose now that the initial level is  close to 
but the final quantum number $m$ far away from $j$ such that $j-m$ is of
the order of $j$. For simplicity we shall also assume that $m$ is not
close to $-j$. In classical terms, we take the Bloch vector as initially
pointing almost to the north pole, but we wait long enough for it to
develop a substantial component transverse to the polar orientation,
i.e. a strong dipole moment; by excluding the late stages of near south
polar orientation we confine ourselves to the phase of brightest
radiation which actually gave rise to the term ``super''radiance.

Under the limitations on $m,n$ just specified the saddle-point equation
\rf{pab131} can still be solved analytically. The important fact is that
the function $a$ takes on values close to unity. More accurately, it can
be shown that the difference $1-a$ is of the same order of magnitude as
\be
\delta_{\nu}\equiv 1-\nu\,,
\ee
the deviation of the initial classical  coordinate from unity.
It will be convenient to introduce the quantum time shift
\be
\tau'=\tau-\tau_{class}(\mu,\nu)\,,
\label{tautauprim}
\ee
where $\tau_{class}(\mu,\nu)$ denotes the classical time of travel from
$\nu$ to $\mu$ given by \rf{pab1311}; in the situation under study it is
\be
\tau_{class}\approx \frac 1 2 \ln \frac 2 {\delta_{\nu}}-\frac 1 2
\ln \frac{1+\mu}{1-\mu}\,.
\label{tauappa}
\ee
We can now write $a$ as
\be
a\approx 1-\lt(1-\E^{-2\tau'}\rt)\delta_{\nu}.
\ee

By similarly evaluating the other ingredients in the propagator
\rf{uniform} to leading order in $\delta_{\nu}$ and in addition
replacing all factorials but $(j-n)!=l!$ by their Stirling estimates we
come to
\be 
D_{mn}= \frac 2 {J\lt(1-\mu^2\rt)}
\frac {\lt(l+\frac 1 2 \rt)^{l+1}}{l\;!}
\exp\lt[-2(l+1)\tau'-\lt(l+\frac 1 2 \rt)\E^{-2\tau'}\rt]_{l=j-n}.
\label{refux}
\ee
To connect with wellknown results we ban the quantum time shift $\tau'$
by substituting \rf{tautauprim}, \rf{tauappa} and introduce the rescaled
variables
\be
z= 2J\E^{-2\tau},\quad
x = z\frac{1-\mu}{1+\mu}\,.
\label{refex}
\ee
The propagator thus assumes the equivalent form
 \be
D_{mn}(\tau)=\lt.\frac 2 {J\lt(1-\mu^2\rt)}
\frac {x^{l+1}\E^{-x}}{l!}\rt|_{l=j-n,\;\mu=(m-1)/J}.
\label{propup}
\ee
The special case of full initial excitation, $l=j-n=0$, yields a
distribution first derived by De Giorgio and Ghielmetti
\cite{Giorgio1,Giorgio2}.

Contact with several previous treatments of superradiance is made by
considering the bright-stage propagator \rf{propup} for high initial
excitation as a function $D(\mu,\tau;n)$ of the final coordinate $\mu$
and the time $\tau$ and verifying it to obey the first-order partial
differential equation
\be
\frac{\partial D}{\partial\tau}=\frac{\partial}{\partial\mu}(1-\mu^2)D\,.
\label{Liouville}
\ee
Obviously, that dynamics is devoid of quantum effects: The propagator
$D$ drifts along the characteristics of \ref{Liouville}, i.e. the fully
classical trajectories \ref{pab1311} 
\be
D(\mu,\tau;n)=\frac{1-\nu(\mu,\tau)^2}{1-\mu^2}D(\nu(\mu,\tau),0;n)\,,
\label{silly}
\ee
where $\nu(\mu,\tau)$ is the time reversed classical trajectory
obtained by solving \rf{pab1311} for $\nu$.
All quantum effects inherent in the superradiant pulses then originate
solely from an effective initial distribution $D(\mu,0;n)$ which we read from
\rf{propup} by there setting $\tau=0$,
\be
D(\mu,0;n)=\frac{2}{J(1-\mu^2)(j-n)!}
\lt(2J\frac{1-\mu}{1+\mu}\rt)^{j-n+1}
\exp\left(-2J\frac{1-\mu}{1+\mu}\right)\,.
\ee
We should emphasize that this effective initial distribution does not
coincide with the true sharp initial form of the propagator, simply because
our asymptotic propagator  (\ref{propup}) is not valid at small times. The
essence of the earlier theories of Refs. \cite{Giorgio1,Giorgio2,HG} is thus
recovered: 
Each run of a superradiant decay of a highly excited atomic initial
state produces a macroscopic, i.e. classical radiation pulse originating
from effectively random initial data, the latter reflecting quantum
fluctuations.

\subsection{Time dependent expectation values}

We shall here establish a master formula for the
set of ``moments'' defined as
\be
M_{ks}(\tau;l)=\mbox{tr} \lt[\hat{\rho}(j-l;\tau)J_+^sJ^k_zJ_-^s\rt]
\ee
with nonnegative integers $k,s,l$  and $\hat{\rho}(j-l;\tau)$ the
density operator originating from the pure initial state  $\ket{j,j-l}$.
In the case of $j$ much greater than 1 and $k,s,l$ much smaller than $j$
the average $M_{ks}(\tau,l)$ can be written in the form of an integral over
the classical variable $\mu$ with the propagator $JD_{m,j-l}(\tau)\equiv
D(\mu,\nu,\tau)$ as a weight,
\be
M_{ks}(\tau;l)=J^{2s+k}\int_{-1}^1 D(\mu,\nu,\tau)(1-\mu^2)^s \mu^kd\mu.
\label{intup}
\ee
Upon employing the propagator \rf{propup} pertinent to the most highly
excited initial states, changing the integration variable to $x$ [cf.
\rf{refex}], and once more using the rescaled time $z$ from \rf{refex}
we recover \be M_{ks}(\tau;l)=\frac{J^{2s+k} (4z)^s \E^z}
{l!}\int_{0}^{\infty}\frac {x^{l+s}(z-x)^k}{(z+x)^{2s+k}}\E^{-x}dx \,,
\ee an asymptotic result found by rather different methods in
\cite{HG,GH}. It has a scaling form inasmuch as
$M_{ks}(\tau;l)J^{2s+k}$ depends on $J$ and $\tau$ only through the
single combination $z$.

\section{Passage time distribution}
 
In the classical picture of superradiance the Bloch vector starts its
downward motion from a certain initial angle $\Theta_0$ and crosses the
latitude $\Theta$ at a strictly definite time $\tau_{class}(\mu,\nu)$
with $\nu=\cos \Theta_0,\;\mu=\cos \Theta$. In other words, the
classical probability density of the times of crossing a given
coordinate $\mu$ on the way from the initial point $\nu$ is given by the
delta function $\delta(\tau-\tau_{class}(\mu,\nu))=\delta(\tau')$; the
quantum time shift $\tau'$ defined in \rf{tautauprim} is strictly zero
in the classical limit.

Let us now introduce the quantum mechanical generalization of the
classically sharp passage time distribution. According to the master
equation (\ref{master2}) for the densities, the change of the probability
for the system to be in level $m$ during the time interval $d\tau$ equals
$\lt(g_{m+1}\rho_{m+1}-g_m\rho_m\rt)d\tau$. The quantity $g_m\rho_m(\tau)d\tau$
is obviously the probability for the atoms to go down from level $m$ to
level $m-1$ during the time interval $[\tau,\;\tau+d\tau]$ and

\be P_m(\tau)= g_m\rho_m(\tau) \ee is the corresponding probability density
for the time of passage through level $m$. In particular, by stipulating
the atoms to have started from the pure state $\ket{jn}$ with $n>m$, we
specify the passage time distribution as proportional to the propagator,
\be P_m(\tau;n)=g_m D_{mn}(\tau).
\ee
By simply integrating $\dot{D}_{mn}(\tau)$ as given by the master equation
(\ref{master2}) one easily shows that our passage time distribution is
properly normalized to unity,
\be
\int_0^{\infty} P_m(\tau;n)d\tau=1.
\ee 

Our uniform approximation for the propagator allows to easily and
accurately calculate the passage time distribution.
In particular, if the initial state is not close to the
north pole, the function $P_m(\tau;n)$ is just a somewhat widened
variant of the classical delta distribution, with a width inversely
proportional to the square root of the second derivative
$J\Phi_{\mu\mu}$ at the maximum of the exponent in the semiclassical
approximation \rf{asymd2}.

However, for the more interesting initial states of highest excitation,
the passage time distribution has little in common with its classical
analogue. As follows from \rf{refux} in the case of the initial state
$\ket{j,j-l}$ with $l\ll j$ we rather get
\be
P_m(\tau;j-l)=\frac 2 {l!}\lt(l+\frac 1 2\rt)^{l+1}
\exp\lt[ -2(l+1)\tau'-\lt(l+\frac 1 2\rt)\E^{-2\tau'}\rt].
\label{timdisel}
\ee
This density depends only on $l$ and $\tau'$. It gives directly the time
distribution of the $m\to m-1$ transition with respect to the classical
time which corresponds to $\tau'=0$.

The absence of any explicit dependence on $m$ and $j$ means that the
time distributions of probability calculated for different values of
these quantum numbers but the same $l=j-n$ differ only by a trivial time
shift equal to the change in the classical time $\tau_{class}$. In
particular, the standard deviation of the time of crossing the $m$th
level, $\Delta \tau= \sqrt{<\tau^2>-<\tau>^2}$ with
\be
<\tau^k>=\int_0^{\infty}\tau^k P_m(\tau;j-l)d\tau\,,
\label{inteta}\ee
is a function of $l$ only. 
The  integrals \rf{inteta} are easily calculated and give the mean
passage time and the standard deviation as
\bea
<\tau>&=&\tau_{class}+\frac 1 2 \lt[{\bf C}+\ln \lt(l+\frac 1 2 \rt)-\sum_{k=1}^l\frac 1 k \rt],\no\\
\Delta \tau&=&\frac 1 2 \lt(\frac {\pi^2} 6 -\sum_{k=1}^l\frac 1 {k^2}\rt)^{1/2},
\eea
where ${\bf C}=0.5772156649\ldots$ is Euler's constant; in the case $l=0$
the sums over $k$ are absent. 

When $l$ becomes large compared with unity the distribution \rf{timdisel}
becomes sharply peaked around the  point $\tau'=0$ predicted by the
classical theory. However, as long as $l$ remains of order unity or
even becomes zero as for complete initial excitation the passage time
distribution is rather broad: The relative standard deviation
$\Delta\tau/\langle\tau\rangle$ is 
of order $1/\ln j$; the small initial quantum uncertainty of the
polarization $\sin\theta\approx\theta\propto 1/\sqrt{j}$ is
found to be amplified to macroscopic magnitude in the passage time.

\section{Appendix: Uniform and semiclassical approximations for the
propagator of coherences}

The uniform approximation for the dissipative propagator of the
non-diagonal elements ($k\ne 0$) is obtained via the exact relation
\rf{identity}. In the semiclassical approximation Stirling's formula is
also applied in order to replace $Q_{m\pm k},Q_{n\pm k}$ by smooth
functions of macroscopic arguments. We first note the uniform
approximation for the propagator of coherences
\bea 
\lefteqn{D_{mn}^{k}=\frac{\sqrt{Q_{m-k,n-k}Q_{m+k,n+k}} }{(l-1)!}
\lt(\frac{\E l}{J^2}\rt)^{l}\frac{J^{3/2}}{\sqrt{l\,\Xi}}}\no\\
& & \qquad\qquad\lt.\times\exp\lt\{J\lt[\tau(a^2-1+k^2/J^2)-
\sigma(a,\mu,\nu)\rt]\rt\}\rt|_{l=n-m+1}\,.\label{uniformnd}
\eea
The saddle-point parameter $a$ and the functions $\sigma,\Xi$ do not
depend on $k$ and are determined in exactly the same way as for the
density propagator.

Finally, we note the semiclassical approximation extending \rf{asymd2}
to the propagation of coherences. Since there is an additional quantum
number $k$ whose range goes to infinity when $j\to\infty$, a new
macroscopic variable $\eta=k/J$ has to be introduced. It is notationally
convenient to write the previously incurred function $\sigma(a,\mu,\nu)$
with the help of a new auxiliary function
\be
q(x,y)=(x+y)\ln(x+y)-(x-y)\ln(x-y)\,;
\ee
as  $\sigma(a,\mu,\nu)=q(a,\nu)-q(a,\mu)$. Thus equipped we can present
the propagator of the elements of the density matrix  with skewness $k$ as 
\bea
D_{mn}^k&=&\frac{1}{  \sqrt{[1-\mu-\eta)^2][1-(\mu+\eta)^2]}\sqrt{2\pi J\Xi}}\,
\E^{J\Phi'},\no\\
\Phi'&\equiv&\frac 1 2 \lt[q(1,\nu+\eta)-q(1,\mu+\eta)+q(1,\nu-\eta)
-q(1,\mu-\eta)\rt]\no\\
& & \quad-\sigma(a,\mu,\nu)+\tau(a^2-1+\eta^2)\,.
\eea
{\em Acknowledgments:} This work was supported by the
Sonderforschungsbereich 237 
``Unordnung und gro{\ss}e Fluktuationen''. P.B.~is grateful to the Department
of Theoretical Physics for hospitality during his stay in Essen. He also
acknowledges support of RFFI under grant number N96-02-17937. D.B.~would
like to thank P.B.~for hospitality during his stay in St.Petersburg.

\end{document}